\documentclass[aps,epsf,preprint]{revtex4}
\usepackage{epsfig}
\usepackage{amssymb}
\usepackage{epstopdf}
\usepackage{graphicx}% Include figure files

\begin{document}
\draft

\title{Static and Dynamic Anomalies in a Repulsive Spherical Ramp
Liquid: Theory and Simulation}

\author{ Pradeep Kumar$^1$, Sergey V. Buldyrev$^{1,2}$, Francesco
Sciortino$^3$, Emanuela Zaccarelli$^3$, H. Eugene Stanley$^1$}

\address{$^1$Center for Polymer Studies and Department of Physics\\
  Boston University, Boston, MA 02215 USA\\ $^2$Yeshiva
  University, Department of Physics, 500 W 185th Street\\ New York, NY
  10033 USA\\$^3$Dipartimento di Fisica and INFM CRS-SOFT: Complex Dynamics
  in Structured Systems, Universit\`a di Roma La Sapienza, Piazzale Aldo
  Moro 5, I-00185 Rome, Italy}

\date{kbszs.tex ~~~ 3 November 2004}

\begin{abstract}

We compare theoretical and simulation results for static and dynamic
properties for a model of particles interacting via a spherically
symmetric repulsive ramp potential.  The model displays anomalies
similar to those found in liquid water, namely, expansion upon
cooling and an increase of diffusivity upon compression.  In particular, we
calculate the phase diagram from the simulation and successfully compare
it with the phase diagram obtained using the Rogers-Young (RY) closure
for the Ornstein-Zernike equation.  Both the theoretical and the numerical
calculations confirm the presence of a line of isobaric density maxima,
and lines of compressibility minima and maxima. Indirect evidence of a
liquid-liquid critical point is found. Dynamic properties also show
anomalies. Along constant temperature paths, as the density increases,
the dynamics alternates several times between slowing down and speeding
up, and we associate this behavior with the progressive structuring and
de-structuring of the liquid. Finally we confirm that mode coupling
theory successfully predicts the non-monotonic behavior of dynamics and
the presence of multiple glass phases, providing strong evidence that
structure (the only input of mode coupling theory) controls dynamics.

\end{abstract}

\maketitle

\section{Introduction}

Water and some other liquids exhibit anomalous behavior close to their
freezing lines \cite{pabloreview,angellreview}.  Their phase diagrams
have regions characterized by a negative thermal expansion coefficient,
i.e., these liquids expand upon cooling at certain temperatures and
pressures. Besides the density anomaly, such liquids also have other
peculiar thermodynamic and dynamic behaviors \cite{pgdbook}. For
example, the isothermal compressibility increases upon cooling and the
diffusivity increases upon pressurizing. Usually the region of the
diffusion anomaly is wider than the region of the density anomaly, so
that the latter is completely contained in the former
\cite{pablonature}. The anomalous behavior of the thermodynamic properties
of water has been connected to the existence of a hypothetical
liquid-liquid critical point in deeply supercooled states
\cite{poole1,poole2,stanleymish,Sciortino03}. In the case of water, this
critical point is located in an experimentally unaccessible region.
Recently, using the potential energy landscape formalism, it has been
argued \cite{Sciortino03} that under certain assumptions on the
statistical properties of the potential energy landscape, the existence
of a density anomaly must lead to the existence of a liquid-liquid
critical point.

In the case of water, anomalies are thought to be related to the
tetrahedrality of the interparticle potential. On average, each water
molecule has four nearest neighbors connected by hydrogen bonds.
However, tetrahedrality is not a necessary condition for anomalous
behavior and several spherically symmetric potentials that are indeed able to
generate density and/or diffusion anomalies have been proposed
\cite{gaussiancore,Stillinger97,Jagla99,Sadr98,Scala01,%
Buldyrev02,Stell72,franzese}.  Interestingly, such potentials may be
purely repulsive, providing evidence that different microscopic
mechanisms can generate density anomalies. These potentials can be
regarded as the simplest models which yield water-type thermodynamic and
dynamic anomalies and it is important to fully characterize their
thermodynamic and dynamic behavior.  An additional advantage in studying
spherical potentials is that their behavior can also be studied within a
theoretical framework, for their thermodynamic properties can be
calculated using accurate integral equation closures.

Here we study, using extensive molecular dynamics (MD) simulations, a
specific, spherically symmetric repulsive potential introduced by Jagla
\cite{Jagla99,Sadr98,Scala01,Buldyrev02,Stell72} with the aim of fully
characterizing both static and dynamic extreme loci in the
temperature-density plane. We complement the MD study with integral
theory calculations based on the (thermodynamically consistent)
Rogers-Young (RY) closure, which is known to give accurate results for
other repulsive potentials, such as the square shoulder potential
\cite{lang1999} and the star polymer potential
\cite{likos1,likos2,laurati}.  We also compare numerical results in the
low $T$ region with predictions based on the ideal mode-coupling theory
(MCT) \cite{gotze1,gotze2}, which we solve using the RY static
structure factors as input. We find that MCT is able to predict the
non-monotonic behavior of dynamics and the presence of multiple glass
phases, providing further evidence that structure (the only input of
MCT) controls dynamics in this system.

\section{Methods}

\subsection{Discrete Molecular Dynamic Simulations}

We study the linear ramp potential introduced by Jagla
\cite{Jagla99,Sadr98,Scala01,Buldyrev02,Stell72}

\begin{equation}
U(r) = \left\{
\begin{array}{ll}
\infty  & r < \sigma_{0}\\
U_0(\sigma_1-r)/(\sigma_1-\sigma_0) & \sigma_0<r<\sigma_1,\\ 
0  & r > \sigma_{1}\end{array}\right. 
\label{eq:potential}
\end{equation}
focusing on the specific choice of $\lambda \equiv \sigma_1/\sigma_0=
1.76$, which has been studied previously by Jagla \cite{Jagla99}.  Using
Monte Carlo simulations, Jagla showed that a density maximum is found in
the liquid phase.  For this choice of $\lambda$, the potential in
Eq.~(\ref{eq:potential}) is a good candidate for studying the connection
between thermodynamic and dynamic quantities. In order to study the
dynamic properties, we apply the discrete MD method, approximating the
continuous potential by a sequence of step functions with $n$ small
vertical steps,
\begin{equation}
U_n(r) = \left\{
\begin{array}{ll}
\infty & r < \sigma_{0}\\ k\Delta U & \sigma_{1}-(k+\frac{1}{2})\Delta r
< r < \sigma_{1}-(k-\frac{1}{2})\Delta r\\ 0 & r > \sigma_{1} -
\frac{\Delta r}{2}
\end{array}\right. ,
\label{e-step}
\end{equation}
where $\Delta r\equiv(\sigma_1-\sigma_0)/(n+1/2)$, $\Delta U\equiv
U_0/(n+1/2)$, and $k = 1,2,3,\ldots n$.  The unit of length is
$\sigma_0$, while $U_0$ is the unit of energy.  Temperature is measured
in units of energy, i.e., $k_{B} = 1$.  Simulation time is measured in
units of $\sigma_0\sqrt{m/U_0}$, with $m$ as the particle mass, and
pressure in units of $U_0/\sigma_0^3$.  The number density is defined as
$\rho \equiv N/L^3$, where $L$ is the size of the simulation box and $N$ is 
the number of particles.

The standard discrete molecular dynamics algorithm has been implemented
for particles interacting with step potentials
\cite{rapaport,as2004,sfa2002}.  Between collisions, particles move
along straight lines with constant velocities. When the distance between
the particles becomes equal to the distance for which $U(r)$ has a
discontinuity, the velocities of the interacting particles
instantaneously change. The algorithm calculates the shortest collision
time in the system and propagates the trajectories of particles from one
collision to the next. Calculations of the next collision time are
optimized by dividing the system into small sub-systems, so that
collision times are computed only between particles in neighboring
sub-systems.  Since the total energy $E$ is rigorously conserved, it is
best to study the NVE-ensemble in the cubic box of a fixed volume $V=L^3$
with periodic boundary conditions.

We consider $N=1728$ particles in our simulation. For constant
temperature and pressure simulations, the Berendsen thermostat and
barostat are used. Configurations
in the $NVE$ ensemble are saved on a logarithmic spacing for further
processing, after discarding an initial equilibration period for a time
larger than the correlation time at the particular state point
\cite{nota}.  The diffusion coefficient, measured in units of
$\sigma_0\sqrt{U_0/m}$, is calculated as
\begin{equation}
D=\lim_{t\to \infty} 
{\langle\left[{\bf r}(t'+t)-{\bf r}(t')\right]^2 \rangle _{t'} \over 6t},
\end{equation}
where $\langle\ldots\rangle_{t'}$ denotes an average over all particles
and over all $t'$.  The dynamic structure factor for a given vector
${\bf q}$ is defined as
\begin{equation}
S({\bf q},t)\equiv\left\langle\frac{1}{N}\sum_{i,j}^N \exp( i {\bf
q}\cdot[{\bf r}_i(t'+t)-{\bf r}_j(t')])\right\rangle_{t'},
%\sum_j^N \exp(-2\pi {\bf q}\cdot{\bf r}_j(t)/L)\right\rangle_{t'},
\end{equation}  
where $\langle\ldots\rangle_{t'}$ denotes the average over all $t'$ and
$S({\bf q})\equiv S({\bf q},0)$ is the static structure factor.  The
normalized structure factor
\begin{equation}
\phi({\bf q},t)\equiv {S({\bf q},t)\over S({\bf q},0)}
\label{e-phi}
\end{equation} 
is called the density correlator. The isotropy of the liquid allows us
to average $S({\bf q},t)$ over different ${\bf q}$ with the same
modulus.  In the following, we bin together all ${\bf q}$ within a mesh
$\Delta q=\pi/L$.

% P from simulations must be divided by $9.5/10^{-3}=0.0095$ 
% U must be divided by 9.5
% T must be divided by 9.5
% t must be divided by 10/sqrt{9.5}=3.2444284
% D must be divided by 30.82207 and by 6 so that $<\Delta r^2>=6Dt$. 

\subsection{Rogers-Young Closure}

Integral equation theories are powerful tools for studying the structure
and thermodynamic properties of liquids \cite{hansen,caccamo}.  One
assumes a two-body interaction potential for the particles and
introduces the total pair correlation function $h(r)$, related to the
pair distribution function $g(r) \equiv h(r)+1$, and the direct
correlation function $c(r)$. The goal is to solve the self-consistent
Ornstein-Zernike equation,
\begin{equation}\label{eq:oz}
h(r) = c(r)+\rho \int dr' c(|{\bf r}-{\bf r'}|) h(r'),
\end{equation}
where both $c(r)$ and $h(r)$---or, equivalently, $g(r)$---are unknown.
Equation (\ref{eq:oz}) in Fourier space takes the form
\begin{equation}
h(q) = c(q) + \rho c(q)h(q),
\end{equation}
where $c(q)$ is related to $S(q)$ as $c(q)=(1-1/S(q))/\rho$.

According to the particular form of the interaction potential $U(r)$,
one can choose a certain ansatz for $c(r)$, which relates it to the
interaction potential and to $h(r)$ and allows one to solve
Eq.~(\ref{eq:oz}) analytically in some cases \cite{caccamo} and
numerically otherwise. The two frequently employed closures,
Percus-Yevick (PY) or the hypernetted chain (HNC) \cite{hansen}, suffer from being
thermodynamically inconsistent \cite{caccamo}. This means that although
they can provide good estimates of the static structure factor, they
cannot be reliably used for determining the phase diagram of the system.
More sophisticated closures have thus been developed in recent years,
which have a built-in thermodynamic consistency. This is achieved by
introducing an extra parameter in the ansatz, which can then be
determined to fulfill such a condition.

The RY closure \cite{RY} belongs to the thermodynamically consistent
group of closures obtained by appropriately mixing PY and HNC through
the parameter $\zeta$. The ansatz for $c(r)$ in terms of $h(r)$ becomes
\begin{equation}
\label{eq:ry}
c(r) = \exp[-\beta U(r)]\left[1+\frac{\exp[(h(r) - c(r))f(r)]-1}{f(r)}\right] 
- [h(r) - c(r)+1],
\end{equation}
where $f(r)$ is the `mixing function',
\begin{equation}
f(r)\equiv 1 - \exp(-\zeta r).
\end{equation}
For $\zeta=0$, one recovers PY closure, while for
$\zeta\rightarrow\infty$ Eq.~(\ref{eq:ry}) reduces to the HNC condition.

The OZ equation can be solved using Eq.~(\ref{eq:ry}) for a given
value of $\zeta$. The correct solution corresponds to that value of
$\zeta$ for which the compressibilities $K_T$, calculated using the
``virial'' and the ``fluctuations'' routes agree, ensuring
thermodynamic consistency.  This allows us to reliably use the RY
closure not only to calculate the static structure factor, but also
the phase diagram, as we will do in the following. Moreover, RY gives
particularly good results for a purely repulsive potential, such as
the studied ramp potential.  It has already been successfully tested
against simulations for square shoulder potentials \cite{lang1999},
and for both experiments and simulations for the star polymer
effective potential \cite{likos1,likos2,mayer,laurati}.

\subsection{Mode Coupling Theory}

The density correlator defined in Eq.~(\ref{e-phi}) is the fundamental
quantity of interest in MCT, a set of generalized coupled Langevin
equations which can be closed within certain approximations
\cite{gotze1, gotze2}.  Interesting behavior of these observables arise
when the dynamics of the system becomes slower, i.e., when the dynamical
behavior is of the ``supercooled'' type \cite{sspce}.  A typical
two-step relaxation occurs for the density correlators on approaching
dynamical arrest, indicating the emergence of two distinct time scales
in the system's structural relaxation. A first relaxation process, the
$\beta$ relaxation, occurs at short times, and is due to particles
exploring the cages formed by their nearest neighbors. A second
relaxation, the $\alpha$ relaxation, occurs at longer time scales, when
particles are able to escape the cages. MCT predicts the existence of a
glass transition at a characteristic temperature $T_{MCT}$, where the
time scale of this second relaxation $\tau_{\alpha}$ diverges so that
the particles will always remain trapped in their cages.  For $T < T_{MCT}$
the correlators do not relax any more, reaching a finite plateau value at
long $t$, defined as the nonergodicity parameter $f(q) \equiv
\lim_{t\rightarrow \infty} \phi (q,t)$; $f(q)$ jumps discontinuously
from zero to a finite (critical) value $f_c(q)$ at the transition,
signaling the occurrence of an ergodic (fluid) to a nonergodic (glass)
transition.  The transition is kinetic, i.e., nothing happens at the
thermodynamic properties of the system close to $T_{MCT}$.

MCT predictions are often found to be in agreement with
experimental \cite{goetze} and simulation results \cite{silica},
although in real systems the $\alpha$-relaxation time does not diverge,
but only becomes increasingly larger. This is due to the intervention of
other processes, commonly termed ``hopping'' processes, which restore
ergodicity and are not included in the MCT treatment of the ideal glass
transition described above.

In mathematical terms, the non-ergodicity parameters $f(q)$ are the long
time solutions of the MCT equations, i.e.,
\begin{equation}
\label{eq10}
{f(q)\over 1-f(q)}=m(q),
\end{equation}
where the memory kernel $m(q)$ is quadratic in the correlator
\begin{equation}
\label{eq:kernel}
m(q)={{1}\over{2}}\int {{d^{3}k}\over{(2\pi)^{3}}}
{\cal V}( {\bf q},{\bf k})f({k}) f({|{\bf q}-{\bf k}|}),
\end{equation}
where $k=|{\bf k}|$. The vertex functions ${\cal V}$ are the coupling
constants of the theory, which are given only in terms of the static
structure factor and number density of the system,
\begin{eqnarray}
\label{eq-vertex}
{\cal V}( {\bf q},{\bf k}) = \frac{\rho}{ q^{4}}\left[{\bf q} \cdot(
{\bf q} - {\bf k}) ~c({|{\bf q}-{\bf k}|})+ {\bf q}\cdot {\bf
k}~c({k})\right]^{2} \times \nonumber \\ S({q}) S({k}) S({|{\bf
q}-{\bf k}|}).
\end{eqnarray}
Equations (\ref{eq10}) and (\ref{eq:kernel}) define a system of non
linear equations with a trivial solution $f(q)=0$. However, for certain
values of the vertex functions, the solutions have a bifurcation point,
locating the glass transition. At this transition point a solution
$f(q)>0$ emerges.  The time evolution of the density correlators is
found by solving the full MCT equations,
\begin{equation}
\ddot\phi(q,t)+\Omega^2(q)\phi(q,t)+\int_0^t m(q,t-t')
\dot\phi(q,t')\,dt'=0\;,
\label{eq:mct}
\end{equation}
where $\Omega^2(q)\equiv q^2/(\beta m S(q))$ and $m(q,t)$ is the
time-dependent memory kernel
\begin{equation}
\label{eq:kerneltime}
m(q,t)\equiv {{1}\over{2}}\int {{d^{3}k}\over{(2\pi)^{3}}} {\cal V}
({\bf q},{\bf k})\phi({k},t)\phi({|{\bf q}-{\bf k}|},t).
\end{equation}

The two-step relaxation is well described by MCT through an asymptotic
study of the correlators near the ideal glass solutions.  The
$\alpha$-relaxation is effectively described by a stretched exponential,
\begin{equation}
\phi(q,t) = A_q\exp[-(t/\tau_q)^{\beta_q}],
\label{eq:stex}
\end{equation}
where the amplitude $A_q$ determines the plateau value and $\beta_q<1$.

MCT predicts a power-law divergence of the $\alpha$-relaxation time as
well as a power-law decrease of the diffusivity as the system approaches
the ideal glass transition.  In this study, we compare the dynamic
behavior evaluated from the MD simulations with corresponding MCT
predictions.

\section{Results}

\subsection{Dependence on the Number of Steps in the Ramp}

\begin{figure}[htb]
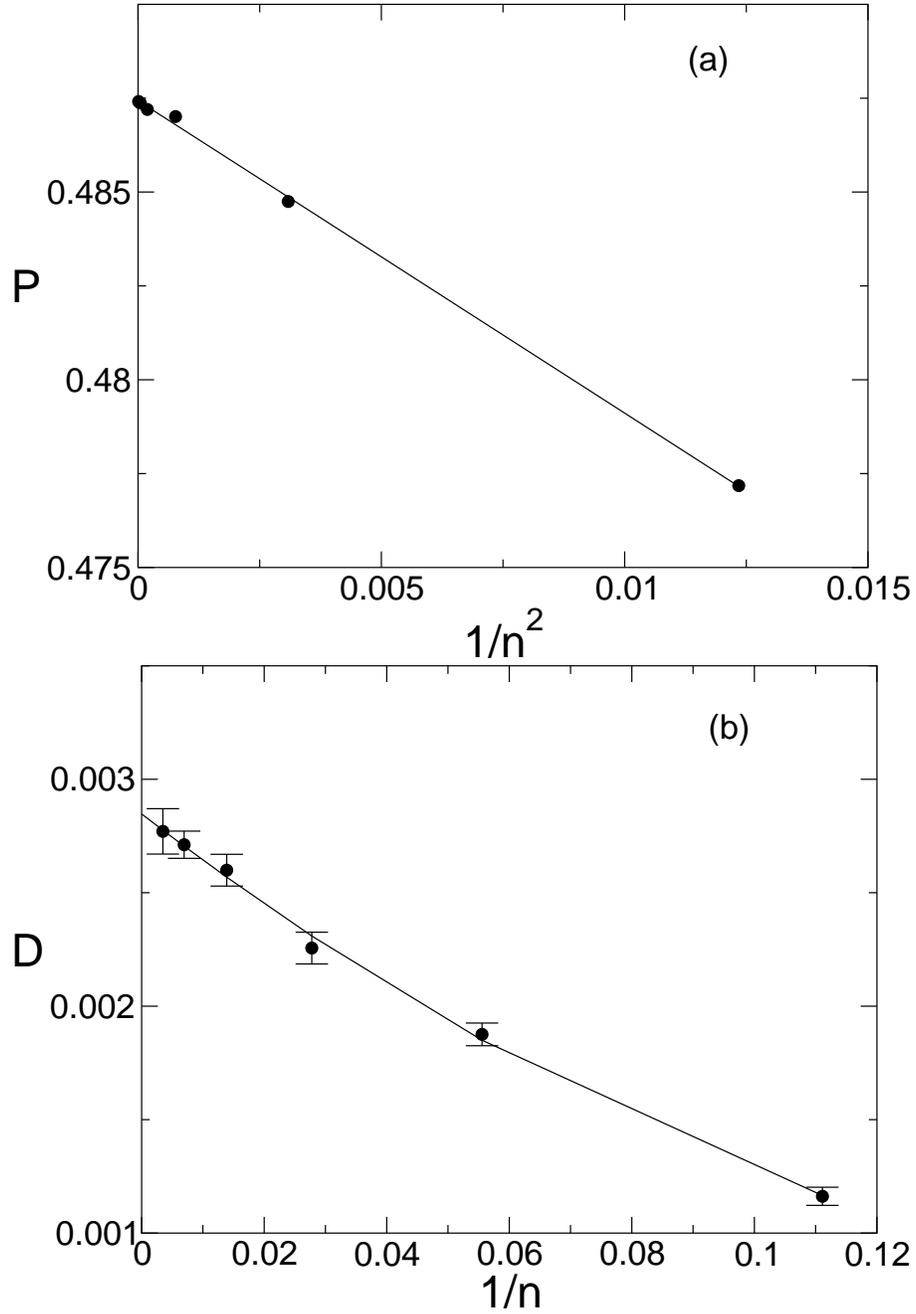

\includegraphics[width=12cm,angle=0]{p-n.eps}
\includegraphics[width=12cm,angle=0]{dc-n.eps}
\caption{(a) Pressure $P$ as a function of $n^{-2}$ [where $n$ is
    defined in Eq.~(\protect\ref{e-step})] for $T= 0.063$ and $\rho =
    0.260$. (b) Diffusion coefficient $D$ as a function of $n^{-1}$ for
    the same state point.}
\label{f-n}
\end{figure}

We first study how the results of molecular dynamic simulations for
the potential $U_n(r)$ in Eq.~(\ref{e-step}) converge to the results for the
continuous ramp potential $U(r)$ in (\ref{eq:potential}), as $n\to\infty$ where
$n$ is the number of approximating steps. The exact equation of state for systems described by a pair potential can be calculated from the
partition function, i.e., by the integral $\int\int...\int
\exp[-\sum_{i>j}U({\bf r}_i-{\bf r}_j)/k_BT]\prod_{i=1}^Nd{\bf r_i}$.
Replacing the continuous potential $U({\bf r}_i-{\bf r}_j)$ by a
step-function is analogous to replacing an integral by a sum in a
rectangular approximation which is known to converge to the integral as
$1/n^2$. Thus we can expect that the pressure $P_n$ obtained in the
discrete molecular dynamics for given $n$ converges to the value for the
continuous potential $P_\infty$ as $1/n^2$.  On the other hand, the
probability for a particle to jump over a step of size $\Delta U$ is
proportional to $\exp(-\Delta U/k_B T) \sim 1 -U_0/n k_B T$. The
diffusion coefficient must be a differentiable function of this
probability. Thus we can expect that the diffusion constant $D_n$ in
discrete MD approaches its limiting value $D_\infty$ as $1/n$. Figures
\ref{f-n}(a) and \ref{f-n}(b) confirm these predictions.  In the
following, we limit ourselves to the case $n=144$ which, as shown in
Figs.~\ref{f-n}(a) and \ref{f-n}(b), is sufficiently close to the $n \to
\infty$ case.

\subsection{Structure Factors and Comparison with RY}

\begin{figure}[htb]
\includegraphics[width=15cm,angle=0]{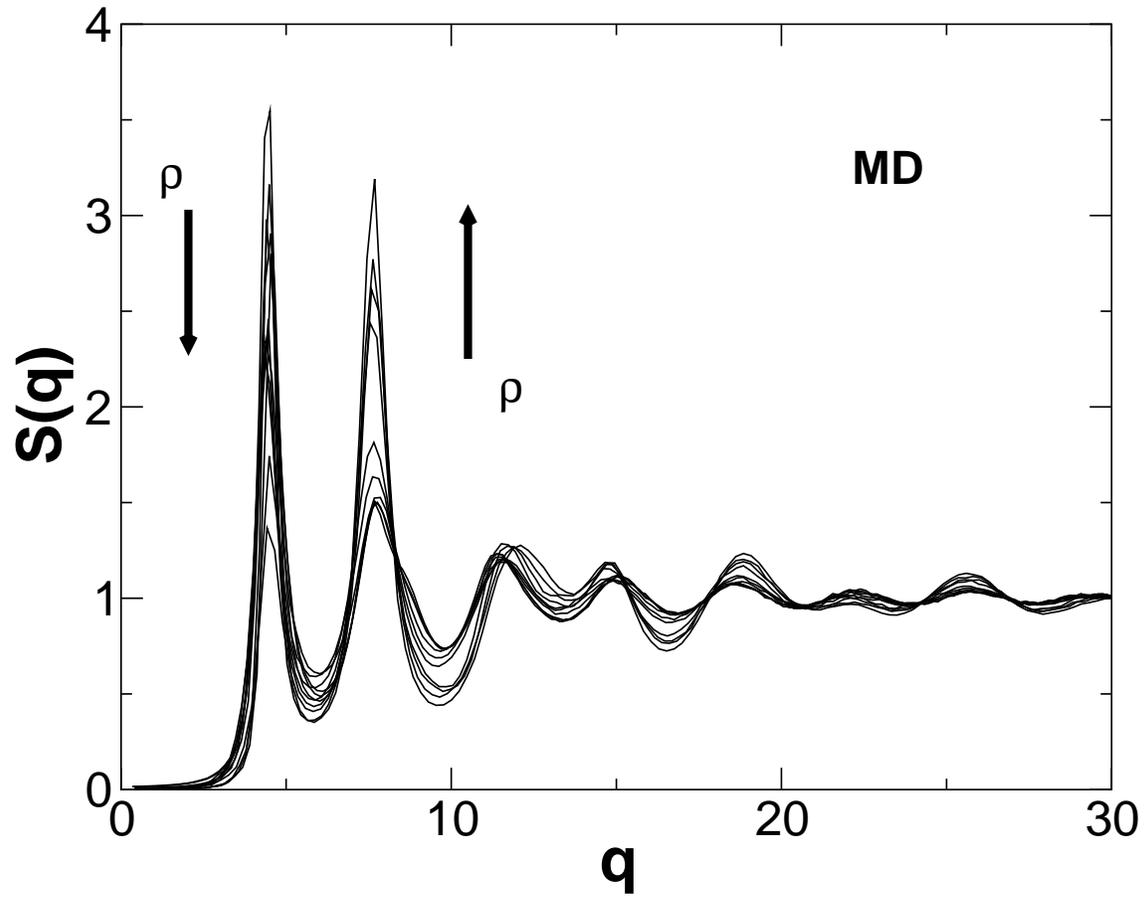} 
\caption{Density dependence of the structure factor at $T=0.063$.  Note
  the progressive reduction of the amplitude of the first peak and the
  progressive increase of the second peak on increasing
  $\rho$. Different curves refer to
  $\rho=0.272$, 0.296, 0.322, 0.352, 0.384, 0.421, 0.464, 0.512, 0.567,
  0.629, and 0.702.}
\label{fig:sqmd}
\end{figure}

Figure~\ref{fig:sqmd} shows the density dependence of the structure
factor, as calculated from the MD at $T=0.063$.  At this low $T$ the
liquid, even at low densities, is significantly structured, as shown
by the large amplitude of the first peak.  On increasing the density,
contrary to the normal liquid behavior, the amplitude of the first
peak is reduced. As discussed below, the destructuring of the liquid
is associated to a speed-up of the particle dynamics. If $\rho$ is
further increased, the second peak becomes dominant and (as discussed
in the following) its increase correlates with a slowing down of the
dynamics.  The first peak is significantly reduced and acts as a
pre-peak on the major peak.  The crossover of the leading amplitude
from the first to the second peak resembles the behavior previously
discovered in star-polymers of large functionality
\cite{anomalousq}. In these systems, the star-star interaction can be
effectively reduced to an ultra-soft repulsive logarithmic interaction
\cite{likos1}. However, the peak positions of $S(q)$ change positions
with density in this case, as opposed to what is found in ramp
potential.

\begin{figure}[htb]
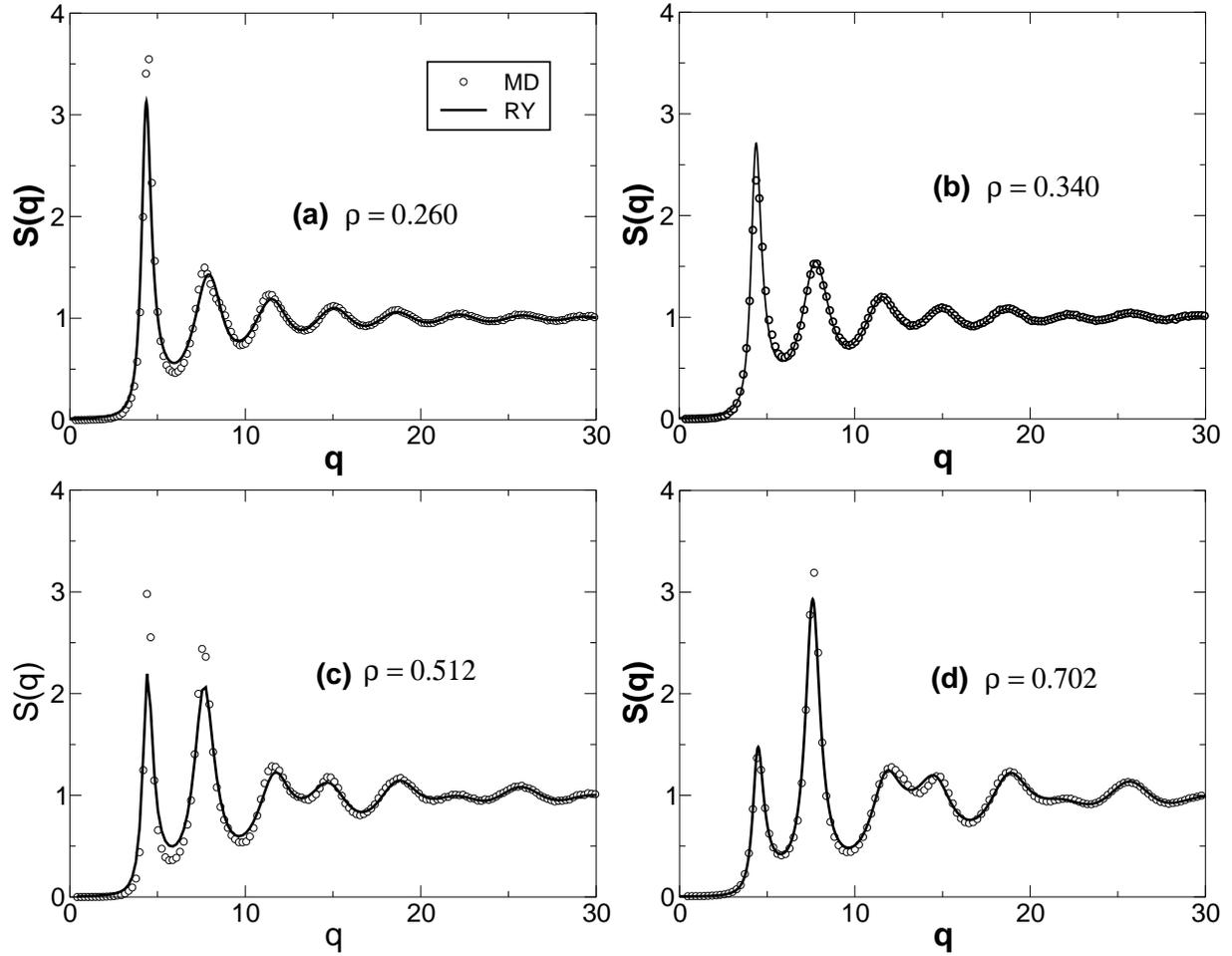

\includegraphics[width=8cm,angle=0]{sqT06L188.eps} 
\includegraphics[width=8cm,angle=0]{sqT06L172.eps}
\pagebreak
\includegraphics[width=8cm,angle=0]{sqT06L150.eps}
\includegraphics[width=8cm,angle=0]{sqT06L135.eps}
\caption{Comparison of the static structure factor, obtained in
simulations and in the RY closure for (a) low density $\rho=0.260$,
(b)--(c) intermediate densities $\rho=0.340$ and $\rho=0.502$, and (d)
high density $\rho=0.702$.}
\label{f-sq}
\end{figure}

Figures \ref{f-sq}(a)--\ref{f-sq}(d) compare the static structure
factor calculated using simulation data along with the results of the
RY-closure for several densities at $T=0.063$. A tolerance of $5.10^{-5}$
is used for the thermodynamic consistency. The RY integral theory
provides a correct description of the $\rho$ dependence of the
dynamics. RY correctly predicts the structure factor at high
densities, while for intermediate and low densities RY gives a fairly
good estimate of the structure factor except that the first peak in
the structure factor is lower than the one calculated from
simulation. RY thus tends to underestimate the structure of the
liquid.

\subsection{Phase Diagram}

\begin{figure}[htb]
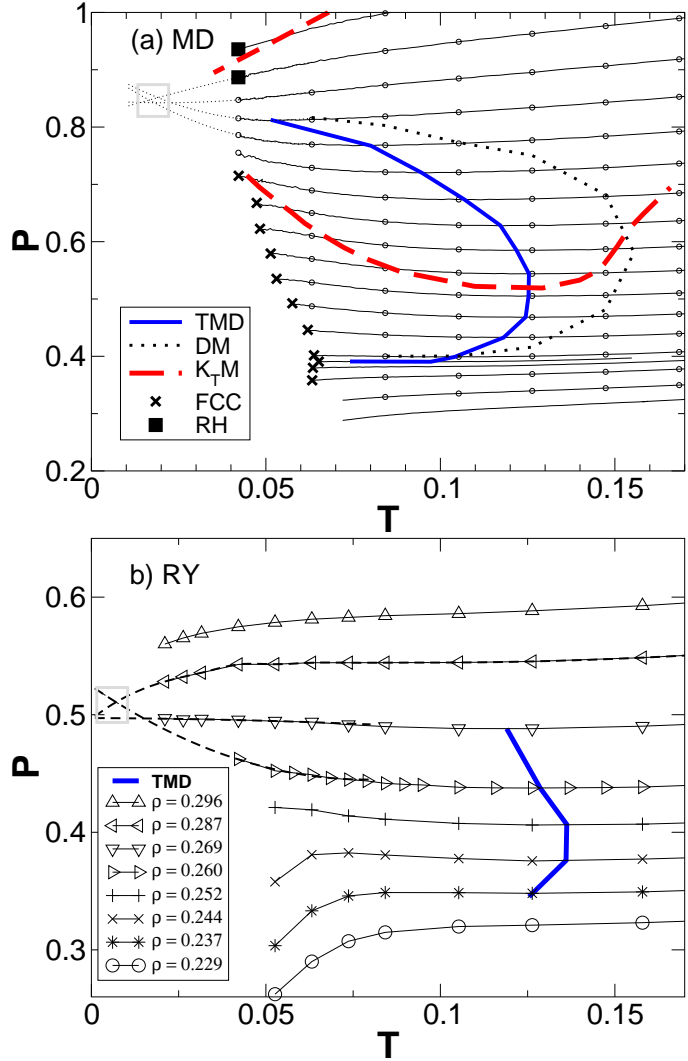

\includegraphics[width=9.0cm,angle=0]{t-p2.eps}
\includegraphics[width=9.0cm, angle=0]{p-ta.eps} 
\caption{(a) Phase diagram of a system of particles interacting via the
  potential defined in Eq.~(\protect\ref{e-step}) with $n=144$.  Lines
  indicate $P(T)$ for several isochores at the following values of
  $\rho$, from top to bottom: 0.378, 0.364, 0.352, 0.340, 0.328, 0.317,
  0.306, 0.296, 0.287, 0.277, 0.269, 0.260, 0.252, 0.244, 0.242, 0.240,
  0.237, 0.229, 0.223.  Also shown are the locus of density maxima
  ($\partial V/\partial T|_{P}=0$) (bold blue line), the locus of compressibility maxima
  and minima $\partial V/\partial P|_T=0$ (dashed red line), and the locus $\partial
  D/\partial P|_T=0$ (bold dotted line). At low $\rho$, lines terminate when the system
  crystallizes or does not equilibrate within the available simulation
  time. Note that different crystals, fcc and rhombohedral, are found
  for $\rho < 0.317 $ and $\rho > 0.352$, respectively.  The
  extrapolated isochores at large $\rho$ cross at a finite $T$ (open
  box), consistent with the possibility of a liquid-liquid critical
  point for $T$ below the ones that can be investigated numerically.
  (b) Corresponding phase diagram obtained within the RY closure. An
  extrapolation of the isochores also shows the possibility of a very
  low temperature liquid-liquid phase transition.}
\label{f-p-t}
\end{figure}

Figure \ref{f-p-t} shows the phase diagram of the model obtained by slow
cooling (thin lines representing isochores) and by accurate simulations
of individual state points (circles).  The simulations for each density
and temperature are initialized with random configurations and are
equilibrated to the desired temperature using Berendsen's thermostat for
a sufficient period of time. The equilibration time was estimated as the
time when the density correlator $\phi({\bf q},t)$ at the first peak of
$S(q)$ decays to zero. After equilibration, each configuration was left
to run at constant energy for a time dependent on the speed of the
dynamics, covering at least ten equilibration times.  The temperature of
maximal density (TMD), shown by a bold line, was obtained by connecting
the points on each isochore corresponding to the minimal pressure, since
these points correspond to the points of minimal volume due to the
general relation
\begin{equation}
\left({\partial V \over \partial T} \right)_P=
-\left({\partial P \over \partial T} \right)_V
\left({\partial V \over \partial P} \right)_T
\end{equation}

and $(\partial V / \partial P)_T<0$ for a mechanically stable system.
Thus, volume increases upon cooling at constant pressure in the region
to the left of the TMD line.  

At low $T$, the system spontaneously crystallizes during the time of the
simulation in different crystalline forms. Consistent with Jagla's
calculation (\cite{Jagla99}), we find that:

\begin{itemize}

\item[{(i)}] At low densities, the system crystallizes into a face
     centered cubic crystal structure upon slow cooling, when the
     temperature reaches the values indicated by crosses. The
     crystallization is marked by a sharp drop in the potential energy,
     associated with a fast release of latent heat. The line connecting
     these points can be regarded as a line of homogeneous nucleation.
     It has a marked negative slope, corresponding to the smaller
     density of the crystalline phase compared to liquid.  The
     equilibrium melting line, which also has negative slope, is
     located at much higher temperatures, so that a large portion of the
     density anomaly lies in the supercooled state. This situation is
     completely analogous to the situation in water, in which the region
     of the density anomaly is located near the negatively-sloped
     freezing line. In the ramp model, the anomaly does not exist below
     $\rho=0.2432$, where the line of the {\it minimal\/} density merges
     with the line of the {\it maximal\/} density.  Interestingly, the
     slope of the homogeneous nucleation line becomes positive below
     this density, as in a normal liquid where the crystal phase has
     a larger density than in the liquid phase.  This behavior may exist
     in stretched water at negative pressures but has never been
     observed experimentally or in simulations.

\item[{(ii)}] In the region of intermediate densities $0.317<\rho
     <0.352$ the liquid does not crystallize and enters the glassy
     state below $T<0.042$.  We are not able to equilibrate the system
     below this temperature.

\item[{(iii)}] For $\rho>0.352$, the system crystallizes into a
     rhombohedral crystal after being equilibrated for a long time at
     $T=0.041$. Interestingly, the density anomaly vanishes at
     densities slightly lower than the density at which this new
     crystal phase emerges. This crystalline phase is characterized by
     parallel columns formed by equidistantly spaced atoms. The
     spacing among these atoms is slightly larger than the hard core
     distance. Thus in this crystalline phase each atom has two
     neighbors in its repulsive ramp.  The projection of these columns
     onto a perpendicular plane forms a triangular lattice with
     spacing approximately equal to the diameter of the repulsive
     ramp. The columns are shifted with respect to each other by one
     third of the nearest-neighbor spacing so that the atoms form
     three crystalline planes, perpendicular to the columns. In each
     of these planes atoms form a triangular lattice with a spacing
     $\sqrt{3}$ times larger than the spacing in the triangular
     lattice formed by the projection of the columns.

\item[{(iv)}] A third distinct (hexagonal) crystalline phase is observed
     for $\rho \approx 0.7865$ (outside the range of densities explored
     in Fig.~\ref{f-p-t}). At this state point the liquid, after some
     initial equilibration, crystallizes into a crystalline phase
     characterized by a hexagonal symmetry of one of its crystalline
     planes.  This crystalline type is also formed by parallel columns
     consisting of equidistantly spaced atoms. The spacing among these
     atoms is slightly larger than the hard core distance. The projection of
     these columns onto a perpendicular plane forms a hexagonal lattice
     with spacing slightly smaller than the hard core. This spacing is
     equal to $\sqrt{3}/2$ of the distance between atoms in the
     columns. The neighboring columns are shifted with respect to each
     other by one half of the atom spacing in the columns, so that the atom
     and its two neighbors in the neighboring column form an equilateral
     triangle. Thus in this crystalline type, each atom has eight
     nearest neighbors in its repulsive ramp: two in the same column and
     six in the three neighboring columns.

\item[{(v)}] For even larger densities not studied in this work,
     hexagonal close packed and finally hard-sphere face cubic
     centered crystals are expected \cite{Jagla99}.

\end{itemize}

We now discuss the possibility of a liquid-liquid critical point in this
model. By quadratically extrapolating the isochores into the glassy state,
we observe a crossing at very low temperatures. This crossing of the
near density isochores is equivalent to $\partial P/ \partial V|_T =0$
and hence predicts the existence of a critical point with coordinates $T
\approx 0.025$, $P \approx 0.838$, and $\rho \approx 0.346$, close to the
largest density at which an isochore develops a negative
slope. Interestingly, the TMD line, if extrapolated, appears to go
directly to this putative critical point.

Another characteristic feature is the behavior of the isothermal
compressibility, which shows an anomalous increase upon cooling between
the lines of maximal and minimal compressibility, are shown in red in
Fig.~\ref{f-p-t}(a). The part of the low density branch with a positive slope
corresponds to a compressibility minima, while the high density branch corresponds to the compressibility maxima. As in water and other
materials, the isothermal compressibility line crosses the TMD line at
the point of its maximal temperature (vertical slope) due to the
mathematical properties of the second derivative of the equation of
state \cite{roma}.  Again, as in water, the line of compressibility
maxima, if extrapolated, seems to approach the putative critical point.

Since the RY closure is thermodynamically consistent, it is possible to
evaluate the phase diagram theoretically as well.  Indeed, once the
static structure factor $S(k)$ is obtained, the pair correlation
function $g(r)$ can be calculated by taking the inverse Fourier
transform of $S(k)$. All other thermodynamic quantities are calculated
using the inter-particle potential function $U(r)$ and $g(r)$. In
Fig.~\ref{f-p-t}(b) we show the phase-diagram consisting of eight
isochores of the system calculated using RY. Comparing
Fig.~\ref{f-p-t}(a) and Fig.~\ref{f-p-t}(b), we can see that the RY predicts
a smaller value of the pressure and also that the density maxima points
are shifted to higher volumes. Despite this discrepancy in $P$, the
shape of the isochores is very well reproduced. Even in the RY case,
density maxima line appears in the phase diagram.  While in the
simulation, extremely slow dynamics prevents access to the region where the
critical point is probably located, in the RY calculations, no
convergence of the parameter $\zeta$ is achieved in the same
region. Again, a smooth extrapolation of the calculated isochores is
consistent with a crossing point, and hence a critical point.

The thermodynamic behavior discussed above is analogous to the behavior
of the one-dimensional model for which an exact solution can be obtained
(see Appendix).

\subsection{Dynamics}

Next we focus on the dynamic properties of the
model. Figure~\ref{f-p-t}(a) shows the lines of diffusivity maxima and
minima, i.e., the locus of points where $\partial D/\partial P|_T=0$ in
the phase diagram.  The region of the anomalous increase of $D$ upon
compression ($\partial D/\partial P|_T>0$) embraces the region of
density anomaly as it does in two-dimensional models as well as in
water \cite{sfa2002,pablonature}.

\begin{figure}[htb]
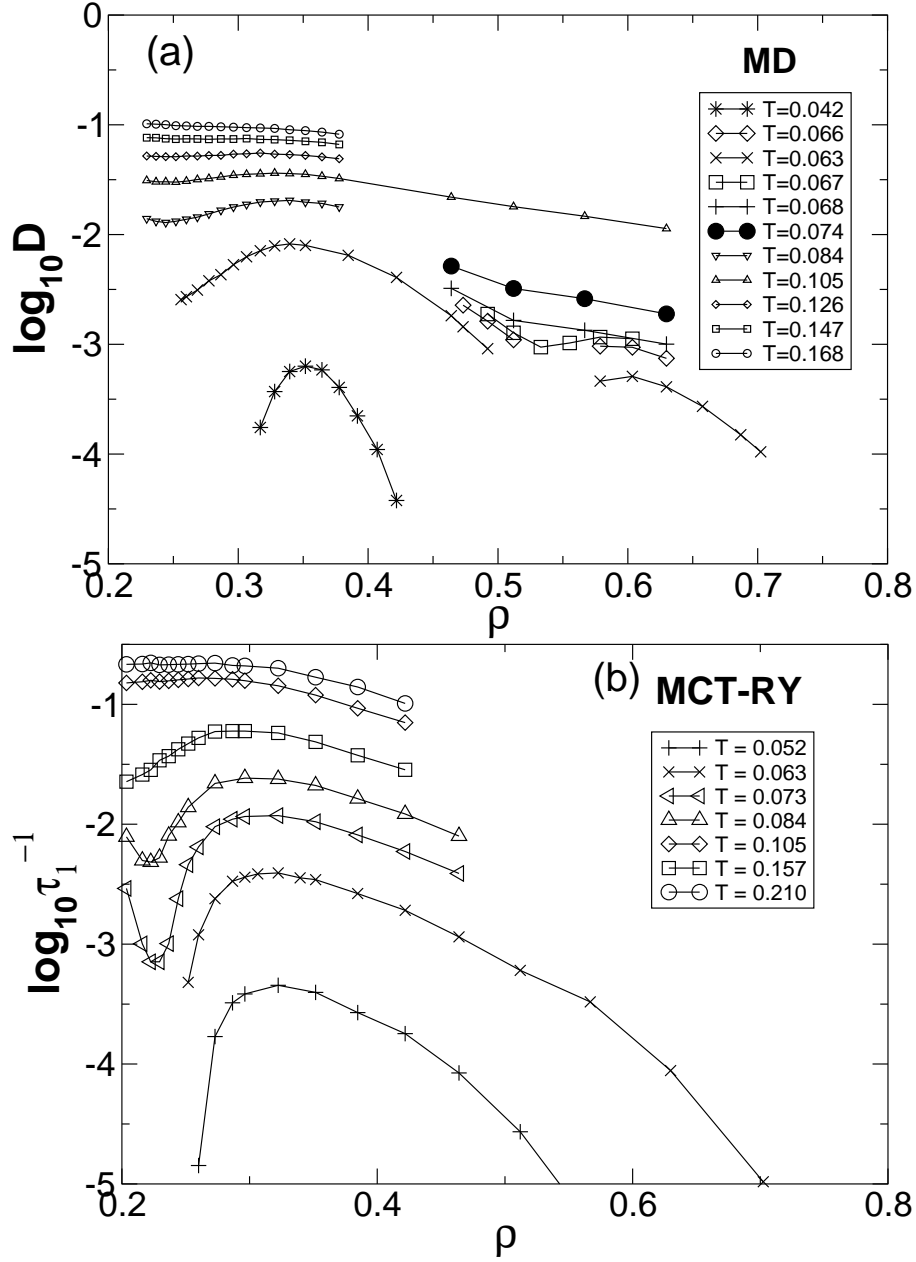

\includegraphics[width=12cm,angle=0]{dc-new.eps}
\includegraphics[width=12cm,angle=0]{tau.eps} 
\caption{(a) The behavior of the diffusion coefficient as a function of
density $\rho$ for several isotherms obtained from MD simulations. (b)
The density autocorrelation time $\tau_1^{-1}$ from MCT-RY.  At low
temperatures when the density is very high or low, $\tau_1$ increases
sharply, showing the transition to a glassy state. Note the similar
behavior of $D$ and $\tau_1^{-1}$. }
\label{f-dc}
\end{figure}

Figure \ref{f-dc}(a) shows the density dependence of $D$ along several
isotherms in a density range extending to $\rho=0.7$.  While only one
minimum and one maximum are observed at high $T$, at low $T$, data
suggest the possibility of a more complex behavior of the density
dependence of $D$. Indeed, for the range of densities
$0.492<\rho<0.579$, the diffusion coefficient drops sharply for
$T<0.067$, and then recovers at higher densities, suggesting the
presence of more than one locus of diffusion maxima. It is interesting
to observe that this second maximum separates regions with different
underlying crystal structures (orthorhombic and hexagonal)
\cite{checkx}, similar to the location of the low $\rho$ maximum,
separating regions where the fcc and orthorhombic crystals are
respectively stable.

It is interesting to compare the $\rho$ dependence of the characteristic
times numerically calculated with the prediction of mode coupling theory
using RY results for $S(q)$ (MCT-RY). A comparison with the dynamics
predicted by MCT can be performed by comparing the $T$ and $\rho$
dependence of $D$ with the $T$ and $\rho$ dependence of the (inverse)
characteristic time scale of decay of density fluctuations. This time
can be calculated within MCT, from the solutions of
Eq.~(\ref{eq:mct}). More specifically, for each wavevector $q$, the
decay time can be defined as the value at which $\phi(q,t)$ reaches the
value $1/e$.  Figure~\ref{f-dc}(b) shows the inverse relaxation time at
the wavevector corresponding to the first peak of the structure factor
$\tau_1^{-1}$ as a function of $\rho$ for different $T$.  The same
sequence of minima and maxima characteristic of the MD diffusivity data
is also found in the prediction of MCT-RY.  This agreement is consistent
with the possibility that the structure factor, the only input in the
MCT, also encodes the system's dynamic properties.

\begin{figure}[htb]
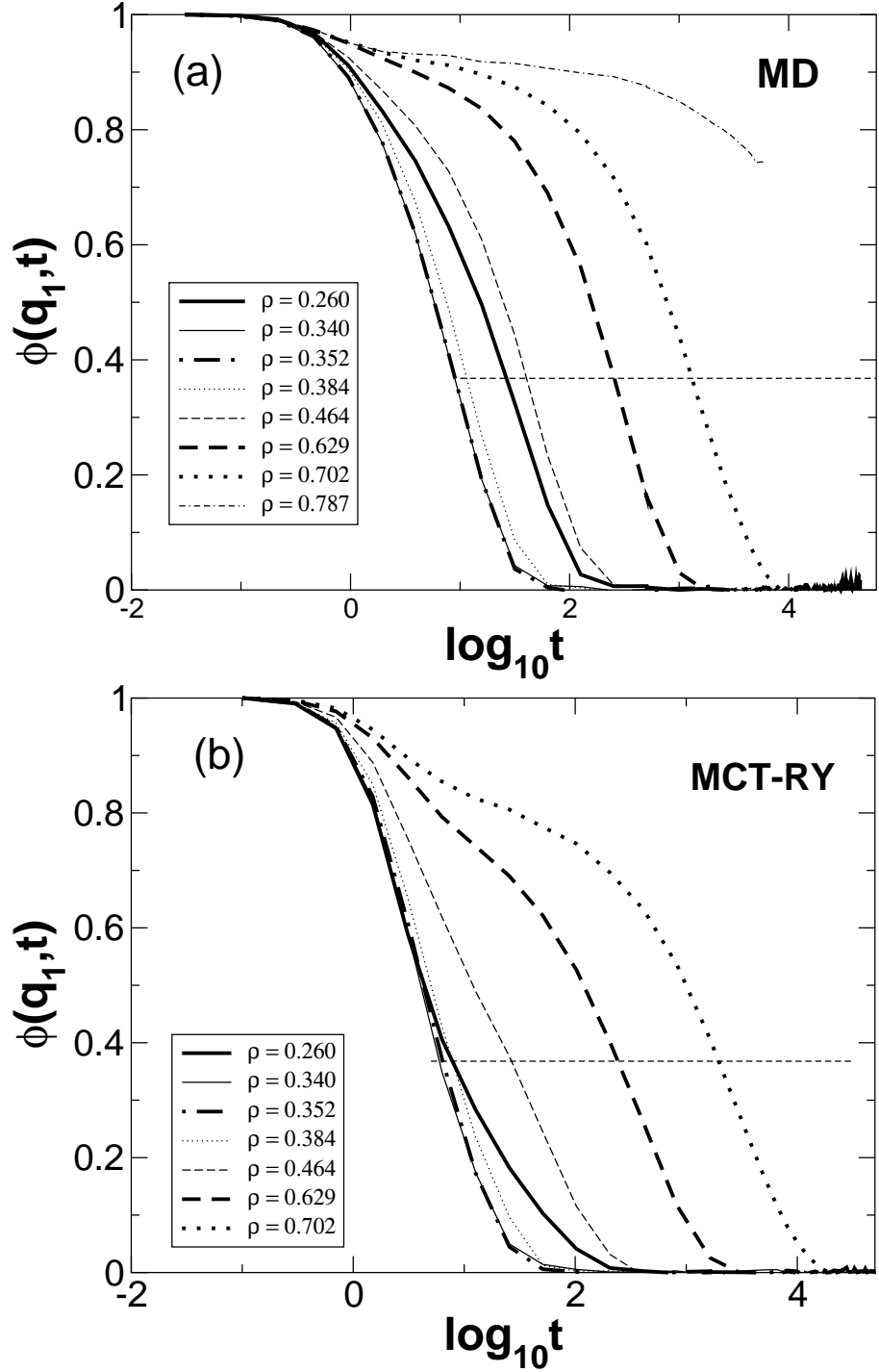

\includegraphics[width=12cm,angle=0]{all.eps}
\includegraphics[width=12cm,angle=0]{all-RYT06.eps}
\caption{The behavior of the correlators for the wave vector $q_1$
   corresponding to the first peak of the static structure factor for
   $T=0.063$ for (a) MD and (b) MCT-RY. The dashed horizontal line
   indicates the $1/e$ value used to define the characteristic time
   $\tau_1$.}
\label{fig:phit}
\end{figure}

To better characterize the low temperature dynamics and investigate the
possibility of different glasses, we next study the decay of the density
autocorrelation functions $\phi(q,t)$ as a function of $t$ at
$T=0.063$ in Fig.~\ref{fig:phit}, and calculate $\phi(q,t)$ from the MD
trajectories Fig.~\ref{fig:phit}(a). The non-monotonic behavior of $D$
and $\tau_1$ discussed in the previous figure is also seen in the decay
of $\phi(q,t)$.  At the highest studied density $\rho>0.787$, the
correlator does not decay to zero in the time scale of our
simulations. For lower $\phi$, the decay becomes faster until
$\phi=0.352$ and then it starts to slow down again.  For $\phi < 0.26$,
crystallization prevents the approach to a glass state.  The decay of
the correlation functions (at $q$ corresponding to the first peak of the
structure factor) can be well represented via a stretched exponential
decay [Eq.~(\ref{eq:stex})] at very high densities, $\rho=0.702$
($\beta=0.89$), and by simple exponential decay ($\beta=1$) at low
$\rho$.  A similar behavior is seen in Fig.~\ref{fig:phit}(b) where the
predictions of the MCT equations are shown. Again, on decreasing $\rho$,
first a speed-up and then a slowing down of the dynamics is observed.

\begin{figure}[htb]
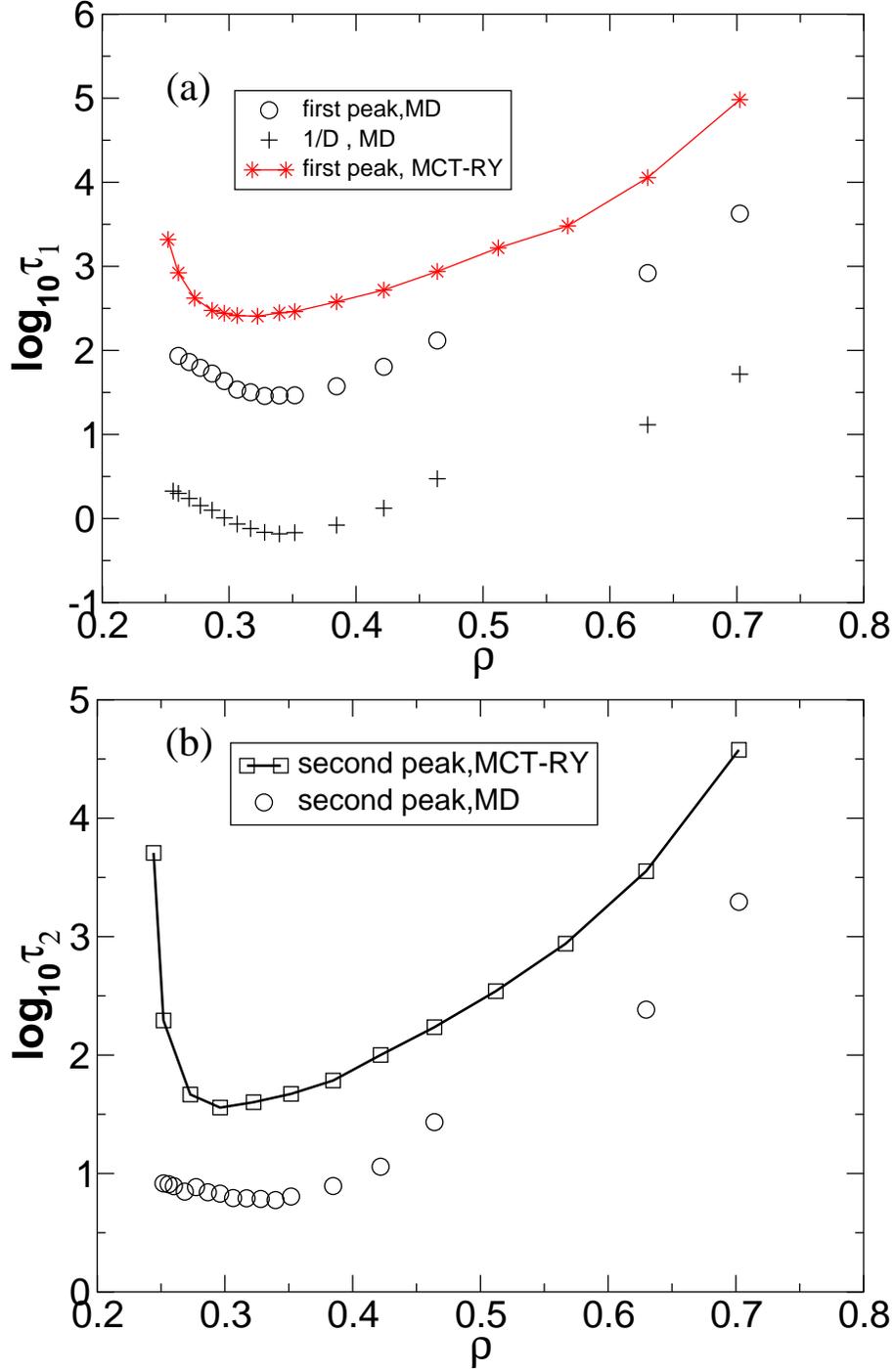

\includegraphics[width=12cm,angle=0]{tau-DMDa.eps}
\includegraphics[width=12cm,angle=0]{tau-DMDb.eps}
\caption{The density behavior of the correlation times (a) $\tau_1$, and
   (b) $\tau_2$, obtained directly from MD simulations for $T=0.063$ and
   for the MCT calculations based on RY. The behavior of the diffusion
   coefficient is also shown. All curves show the anomalous decrease
   with density for small densities.}
\label{fig:tau12}
\end{figure}

To quantify the comparison between MD and MCT-RY, we show the density
dependence of $\tau_1$ and $\tau_2$ in Fig.~\ref{fig:tau12} (the
characteristic time of the first and second peaks of the structure
factor respectively) both from MD and MCT-RY. The two sets of data show
the same anomalous behavior, showing the minimum corresponding to the
maximal diffusivity at the same $\rho$.

\begin{figure}[htb]
\includegraphics[width=15cm,angle=0]{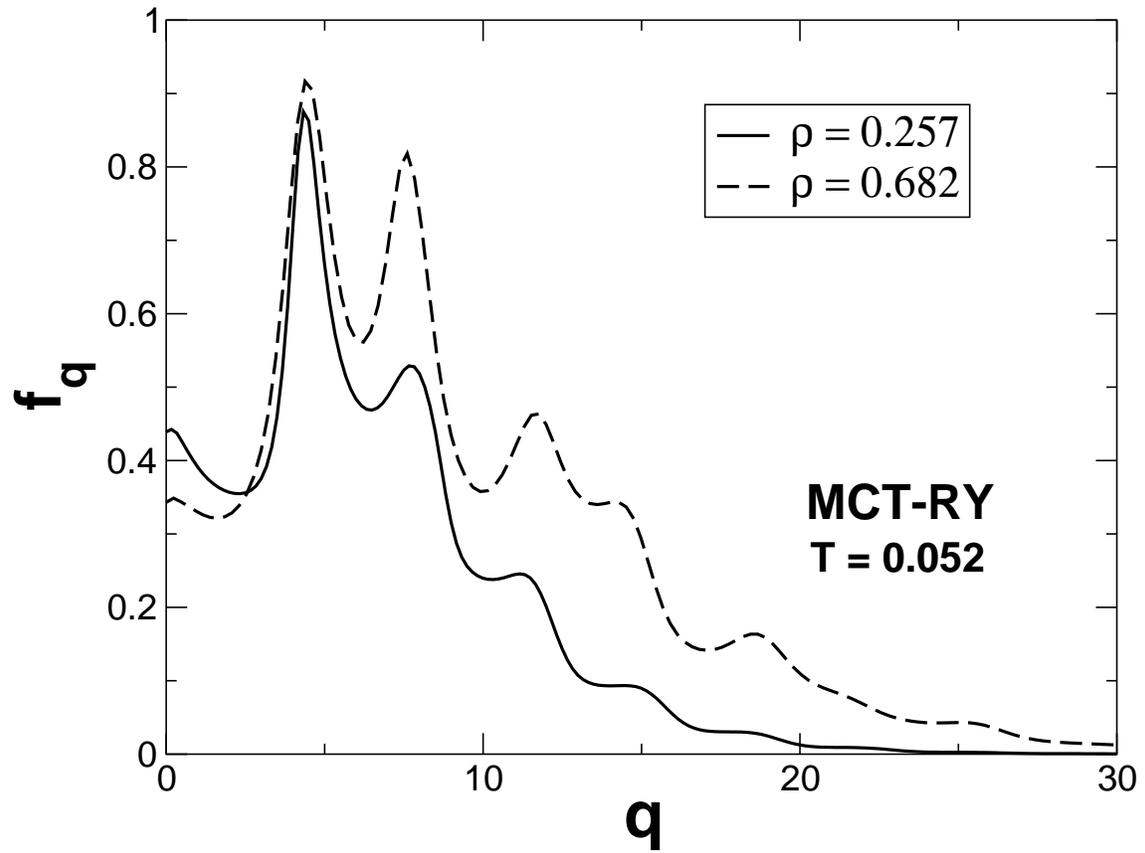} 
\caption{The behavior for $T=0.052$ of the critical non-ergodicity
  parameter $f_q$ for the two critical densities $\rho=0.257$ and
  $\rho=0.682$.}
\label{f-MCT-ne}
\end{figure}

It is interesting to observe that while the slowing down of dynamics can
be numerically followed for a large dynamical range at a high density, a
low density crystallization prevents the generation of a glass
structure. Unlike in MD, RY solutions can be found in a wider density
range. It is thus interesting to ask whether a glass line is predicted
by MCT-RY at low densities and, if so, how the two glasses differ. By
solving the MCT equations for a wider range of densities, we find 
two distinct glasses at $T=0.052$: $\rho=0.682$ and $\rho=0.257$. These two
glasses are characterized by different critical non-ergodicity
parameters (shown in Fig.~\ref{f-MCT-ne}).  While the low density glass
non ergodicity factor is similar to the one found in hard sphere
systems, the high density glass is characterized by large amplitudes
both at the first and second peak of $S(q)$. This resembles the one
found in star polymers at high density, though the amplitude of the
second peak in the latter becomes larger than that of the first peak.

\section{Summary and Conclusions}

In this work we have presented a systematic numerical study of the
static and dynamic properties of a system of particles interacting with
a spherical repulsive potential, with a range of interaction of the
order of the particle diameter. The simplicity of the model makes it
possible to study it with efficient numerical algorithms and,
theoretically, with self-consistent integral theories
\cite{caccamo}. Our results accomplish the following:

\begin{itemize}

\item[{(i)}] They confirm the previous theoretical calculations by
    Jagla~\cite{Jagla99} concerning the existence of a line of density
    maxima in the phase diagram of this potential. Results also provide
    an accurate evaluation of this line as well as of the line of
    compressibility maxima and minima.

\item[{(ii)}] They confirm that different crystal structures are found at low
    temperatures, depending on the density \cite{Jagla99}, and provide
    estimates of the homogeneous nucleation line for the liquid-crystal
    transitions.

\item[{(iii)}] They provide evidence of the possibility of a
    liquid-liquid critical point at $T \approx 0.025$, $P \approx 0.838$ and $\rho \approx
    0.346$. The location of the critical point is below the homogeneous
    nucleation line or glass transition line and cannot be accessed by
    simulations. A theoretical RY calculation of the phase diagram is
    able to reproduce the thermodynamic anomalies. These calculations
    also suggest the possibility of a liquid-liquid critical point but,
    again, its precise location cannot be determined due to the
    impossibility of equating the ``virial'' and ``fluctuation''
    compressibilities with enough accuracy in this region of the phase
    diagram. Thus the existence of the critical point proposed by the
    extrapolation of the equation of state into the deeply supercooled
    region remains questionable \cite{brazillian}.

\item[{(iv)}] They show that the RY closure agrees reasonably well with
    simulations of the system with the repulsive ramp potential and that
    it reproduces the static and dynamic anomalies. The RY closure
    slightly underestimates the pressure, and shifts the anomalies to a
    region of lower density.

\item[{(v)}] They provide evidence that dynamic anomalies in this model
    have a structural origin and they are indeed captured by the MCT-RY
    theory, which uses only structural information as input.  Diffusion
    anomalies are encountered before the density anomalies, consistent
    with the case of water \cite{pablonature}.

\item[{(vi)}] They suggest the possibility of different types of glasses
    in this simple system, consistent with the existence of different
    crystalline phases.  A more extensive study based on the MCT-RY may
    help evaluate the location and the intersections \cite{fabbian99,
    dawson00,zaccarelli02} between different glass transition lines.

\end{itemize}

\subsection*{Acknowledgments}

We thank C. Likos for help with the RY-closure and for extensive
discussions. We also thank NSF Chemistry Grants No.~CHE0096892 and
No.~CHE0404673 and MIUR Cofin 2001 and Firb 2002, and
MRTN-CT-2003-504712 for support.

\appendix
\section{One dimensional system}

Applying the Takahashi method \cite{takahashi}, one can find the
equation of state for the one-dimensional system of particles
interacting via the ramp potential [see Eq.~(\ref{eq:potential})]. In
this case, the partition function $\Psi$ for the Gibbs potential
$G(P,T)$ can be factored, so that
\begin{equation}
G=-\frac{1}{\beta} N\ln \Psi(P,T),
\label{gibbs}
\end{equation}
where  
\begin{equation}
\Psi(P,T)=\int_0^\infty \exp \left( -\beta \left({U(r)+
  Pr}\right)\right) dr. 
\label{Psi}
\end{equation}
Substituting Eq.~(\ref{eq:potential}) into Eq.~(\ref{Psi}) and
integrating, we find that
\begin{equation}
\Psi(P,T)={\exp[-\beta(P\sigma_0+U_0)]-\exp(-\beta P \sigma_1)\over 
\beta (P-P_c)}+{\exp(-\beta \ P\sigma_1)\over \beta P}, 
\end{equation}
where $\beta\equiv 1/k_B T$ and $P_c=U_0/(\sigma_1-\sigma_0)$.
Since $V\equiv(\partial G/\partial P)_T$
\begin{equation}
\rho\equiv N/V=-{\beta\Psi\over(\partial\Psi/\partial P)_T}, 
\end{equation}
which, after differentiation, leads to 
\begin{equation}
\rho=\frac{\frac{e^{-\beta(U_0+\sigma_0 P)}}{(P-P_c)}+e^{-\beta\sigma_1
    P}[\frac{1}{P} -\frac{1}{(P-P_c)}]}{e^{-\beta(U_0+\sigma_0
    P)}[\frac{\sigma_0}{(P-P_c)}+\frac{1}{\beta(P-P_c)^2}]+ e^{-\beta\sigma_1
    P}[\frac{\sigma_1}{P}-\frac{\sigma_1}{(P-P_c)}+\frac{1}{\beta
      P^2}-\frac{1}{\beta(P-P_c)^2}]}
\label{eq:1d}
\end{equation}
Note that $P_c$ plays the role of a critical pressure, at which the
equation of state becomes discontinuous for $T\to 0$. Indeed, 
\begin{equation}
\lim_{T\to 0}\rho=\cases{
1/\sigma_0 & $P>P_c$ \cr
2/(\sigma_1+\sigma_0) & $P=P_c$ \cr
1/\sigma_1 & $P<P_c$}.
\end{equation}
Hence, an infinitesimal increase of pressure around $P_c$ and small $T$
leads to a finite increase of density from $1/\sigma_1$ for $P<P_c$ to
$1/\sigma_0$ for $P>P_c$.  Thus, in the one-dimensional system, there is
an analog of a critical point at $T=T_c=0$, $P=P_c$, and $\rho=\rho_c$,
at which the isothermal compressibility diverges. At $P=P_c$,
Eq.~(\ref{eq:1d}) simplifies to
\begin{equation}
\rho={1+k_BT/U_0\over(\sigma_0+\sigma_1)/2+(k_B T/U_0)\sigma_1+ (k_B
T/U_0)^2(\sigma_1-\sigma_0)}.
\end{equation}
For $P < P_c$ there exists a region in which there is a density anomaly
and a region in which there is a compressibility anomaly, exactly as in
the three-dimensional model studied in this work.

\begin{figure}[htb]
\includegraphics[width=15cm,angle=0]{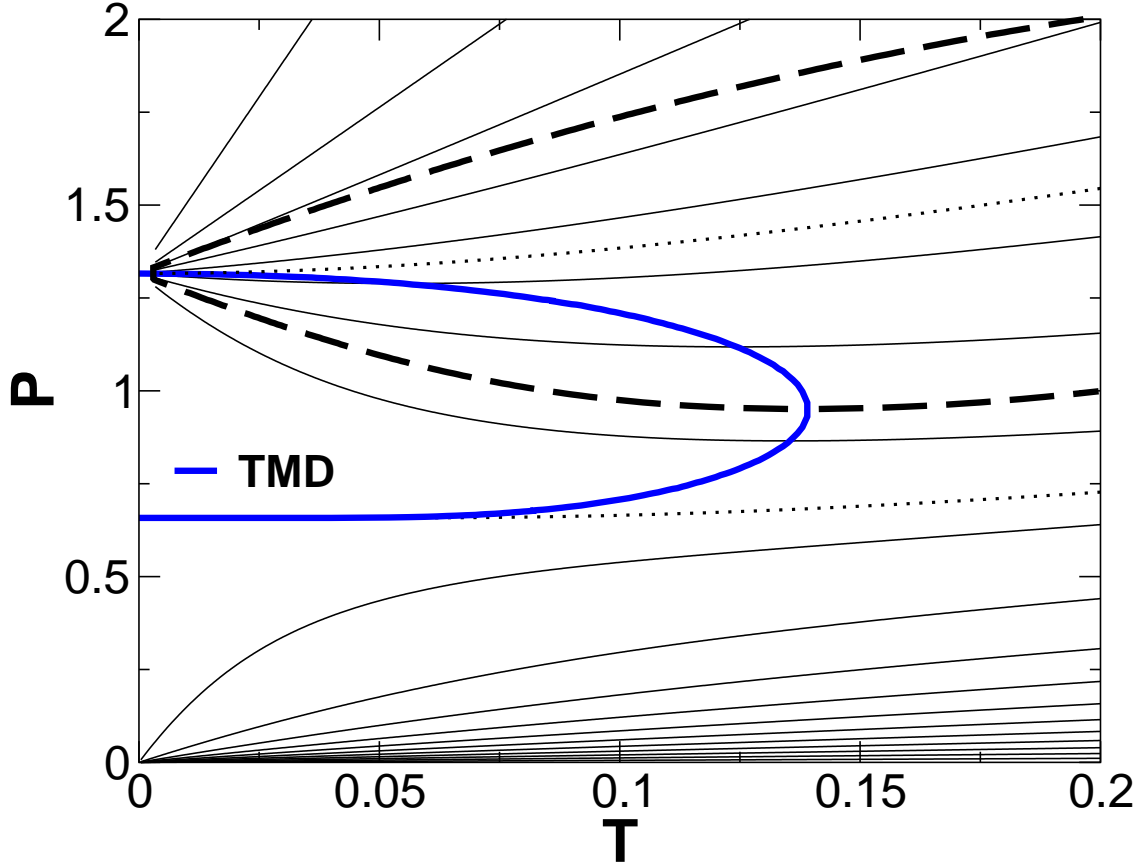}
\caption{The $P-T$ phase diagram of the one-dimensional system with
the ramp potential $(\sigma_0=1, \sigma_1=1.76, U_0=1)$ defined in
Eq.~(\ref{eq:potential}). The thin solid lines are isochores for
$\rho=0.05, 0.1,..., 0.95$ from bottom to top.  The bold dotted lines
are isochores for $\rho=1/\sigma_1$ and $\rho=2/(\sigma_0+\sigma_1)$
between which the density anomaly region bounded by the temperature of the 
maximal density line (bold blue line) is located. The dashed bold line
bounds the region of anomalous isothermal compressibility extrema (both
minima and maxima).}
\label{fig:1d}
\end{figure}

The isochores described by Eq.~(\ref{eq:1d}) are plotted in
Fig.~\ref{fig:1d}.  One can see that the temperature of maximal density
approaches the critical point as $P\to P_c$.

\begin{figure}[htb]
\includegraphics[width=12cm,height=10cm,angle=0]{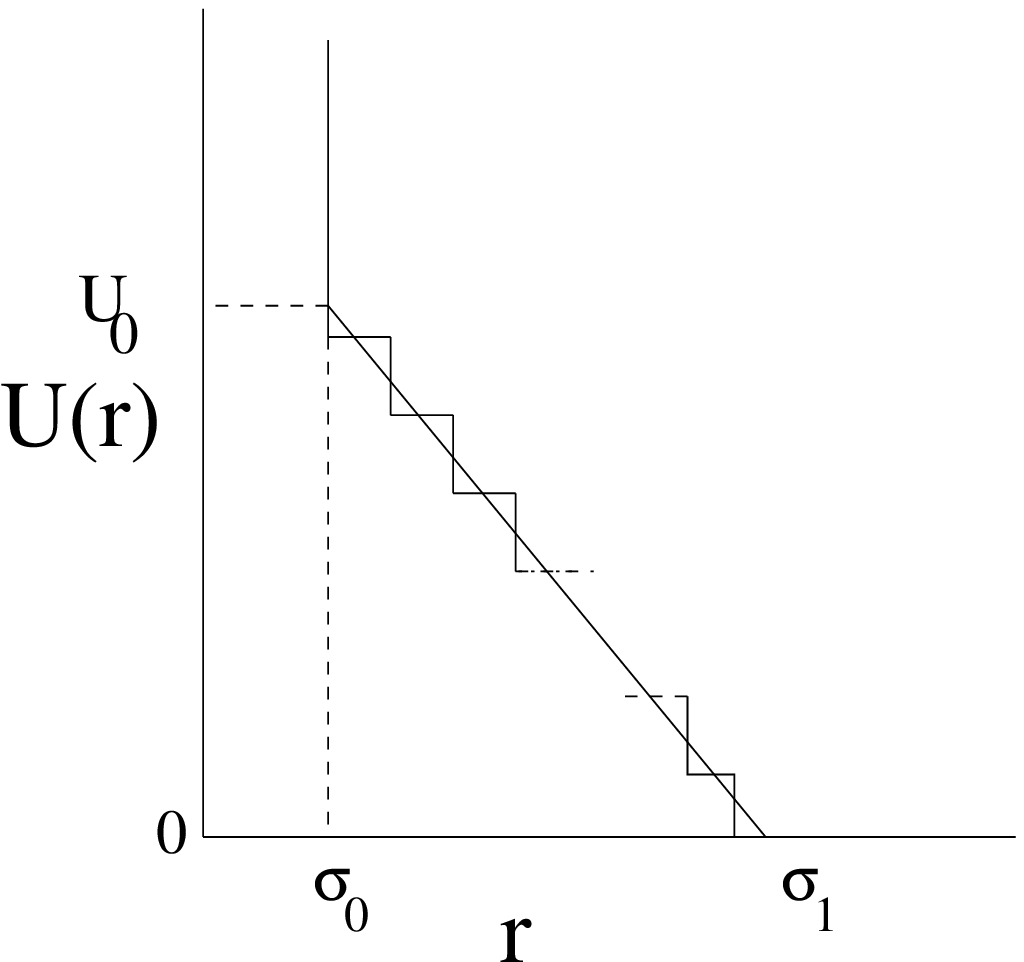}
\caption{Schematic representation of the repulsive ramp potential
(\protect\ref{eq:potential}) and its discontinuous version
(\protect\ref{e-step}).}
\end{figure}

\begin{thebibliography}{99}
\bibitem{nota} Configurations are saved at times $t_{jk} \equiv
(js^m+s^k)\Delta t$, where $k=0,1,..., m$ and $j=0,1, ...,j_{\rm max}$,
For the majority of our state points we use $\Delta t= 0.03082207$,
$m=14$, $j_{\rm max}=100$ and $s=2$.
\bibitem{checkx} One may expect that this part of the phase diagram
   corresponds to partial crystallization.  However, the behavior of the
   system in this region of the phase diagram does not resemble
   crystallization, which at other regions of the phase diagram is
   characterized by a sharp decrease of potential energy with a latent
   heat release. The potential energy at these state points slowly
   decreases with time, resembling the aging of glasses. Also, upon
   heating to a slightly larger temperature, $T=0.0684$ (for which no
   anomalous behavior of diffusion coefficient is observed), the system
   slowly relaxes back to the normal liquid phase. In contrast,
   spontaneous crystallization usually happens in a deeply supercooled
   state so that the true crystal does not melt if the temperature is
   slightly increased.  Also, the structure factors of these phases do
   not develop large, high $q$ peaks, characteristic of the crystalline
   phase.  Thus the nature of these intermediate phases is not clear.
\end{thebibliography}
\end{document}